\documentclass[10pt, conference]{IEEEtran}
\usepackage{amsmath,amssymb,amsfonts,nccmath}
\usepackage{calrsfs}
\DeclareMathAlphabet{\pazocal}{OMS}{zplm}{m}{n}
\usepackage{mathtools} 
\usepackage{algorithmic}
\usepackage{algorithm}
\usepackage[table,xcdraw]{xcolor}
\usepackage{graphicx}
\usepackage{textcomp}
\usepackage[dvipsnames]{xcolor} 
\usepackage[utf8]{inputenc}
\usepackage{textcomp}
\usepackage{longtable}
\usepackage{latexsym}
\usepackage[switch]{lineno}
\usepackage[labelformat=parens]{subfig}
\usepackage[inline]{enumitem}
\usepackage{lipsum}
\usepackage{footnote}
\usepackage{comment}
\usepackage{caption}
\usepackage{fancyhdr}

\definecolor{darkblue}{RGB}{0, 0, 139} 

\def\BibTeX{{\rm B\kern-.05em{\sc i\kern-.025em b}\kern-.08em
    T\kern-.1667em\lower.7ex\hbox{E}\kern-.125emX}}
\usepackage{balance}

\usepackage{booktabs, multirow, siunitx, colortbl, makecell}
\definecolor{Gain}{RGB}{230,255,230}    
\definecolor{Gap}{RGB}{255,230,230}     

\title{Adaptive Intrusion Detection for Evolving RPL IoT Attacks Using Incremental Learning}

\author{%
  \IEEEauthorblockN{%
    Sumeyye Bas\IEEEauthorrefmark{1}\IEEEauthorrefmark{5},
    Kiymet Kaya\IEEEauthorrefmark{1}\IEEEauthorrefmark{4}\IEEEauthorrefmark{6},
    Elif Ak\IEEEauthorrefmark{2},
    Sule Gunduz Oguducu\IEEEauthorrefmark{1}\IEEEauthorrefmark{6},
  }%
  \\
  \IEEEauthorblockA{\IEEEauthorrefmark{1}Istanbul Technical University, Faculty of Computer and Informatics Engineering, Istanbul, Turkey}
      \IEEEauthorblockA{\IEEEauthorrefmark{6}ITU AI Research and 
  Application Center, Istanbul, Türkiye} 
    \IEEEauthorblockA{
    \IEEEauthorrefmark{2}Memorial University, Canada}
      \IEEEauthorblockA{\IEEEauthorrefmark{5}Turkcell, Istanbul, Turkey} 
  \IEEEauthorblockA{\IEEEauthorrefmark{4}BTS Group, Istanbul, Turkey} 

  Email(s): sumeyye.bas@turkcell.com.tr, kayak16@itu.edu.tr, elif.ak@mun.ca, sgunduz@itu.edu.tr}

\begin{document}

\maketitle

\thispagestyle{fancy}   
\fancyhead{}                
\lhead{This paper has been accepted for publication in the IEEE CCNC 2025. This is the author's accepted version of the article. }
\cfoot{}
\renewcommand{\headrulewidth}{0pt}

\begingroup\renewcommand\thefootnote{\textsection}
\endgroup

\begin{abstract}
The routing protocol for low-power and lossy networks (RPL) has become the de facto routing standard for resource-constrained IoT systems, but its lightweight design exposes critical vulnerabilities to a wide range of routing-layer attacks such as hello flood, decreased rank, and version number manipulation. Traditional countermeasures, including protocol-level modifications and machine learning classifiers, can achieve high accuracy against known threats, yet they fail when confronted with novel or zero-day attacks unless fully retrained, an approach that is impractical for dynamic IoT environments. In this paper, we investigate incremental learning as a practical and adaptive strategy for intrusion detection in RPL-based networks. We systematically evaluate five model families, including ensemble models and deep learning models. Our analysis highlights that incremental learning not only restores detection performance on new attack classes but also mitigates catastrophic forgetting of previously learned threats, all while reducing training time compared to full retraining. By combining five diverse models with attack-specific analysis, forgetting behavior, and time efficiency, this study provides systematic evidence that incremental learning offers a scalable pathway to maintain resilient intrusion detection in evolving RPL-based IoT networks.
\end{abstract}


\section{Introduction} \label{sec:intro}





The Internet of Things (IoT) connects many low-power devices, but this increased connectivity also creates new security risks \cite{11045829}. In constrained IoT networks, the routing protocol for low-power and lossy networks (RPL) is a de facto standard for routing \cite{8689098}, used in applications ranging from environmental monitoring to smart healthcare. RPL organizes nodes into a destination oriented directed acyclic graph (DODAG) with a root, using ICMPv6 control messages to establish routes. However, because RPL is designed for efficiency, such as using features like trickle timers and rank metrics, it lacks strong security measures \cite{10379229}, making it vulnerable to various routing attacks such as sinkhole, rank, version number, and flooding-based exploits \cite{10824786}. Recent studies have documented at least fifteen distinct routing attacks capable of exhausting resources, disrupting network topology, and destabilizing entire deployments \cite{10494996}, while zero-day attacks remain an inherent and persistent risk within RPL-based IoT systems. For instance, a newly identified destination advertisement object flooding (DAOF) attack can increase control overhead by more than 65\% and latency by up to 400\% even in modest deployments \cite{10824786}. Such results underline that without \emph{adaptive} defenses, RPL-based IoT networks are at constant risk of disruption, resource exhaustion, and long-term instability.

Traditional countermeasures for RPL have mainly focused on strengthening the protocol itself or using lightweight rule-based mitigation techniques \cite{10494996}. Although these methods can address specific threats, they are reactive, limited in scope, and often require changes to the protocol itself. At the same time, machine learning (ML) methods, including ensemble models (e.g., XGBoost, LightGBM, CatBoost) and deep neural networks (DNN), have demonstrated strong accuracy in identifying known RPL attacks \cite{ref1}. However, these models face a major shortcoming: \textit{when confronted with a new or unseen attack type, their performance decreases unless retrained from scratch}. In dynamic IoT environments, where new attack variants emerge rapidly, retraining is not feasible and is inefficient \cite{app122211598}.


To address this gap, in this paper, we explore a \emph{class incremental learning (CIL)} approach for evolving RPL threats. CIL is a specialized continual fine-tuning process with the aim of handling new classes without forgetting the old ones \cite{Zhou_2024}. Our study targets three common RPL attacks, hello flood (HF), decreased rank (DR), and version number (VN), and evaluates five representative model families used in the current RPL attacks literature: tree ensembles (XGBoost, LightGBM, CatBoost), a DNN, and a graph neural network (GNN). We designed three training regimes that mirror real-world scenarios: i) conventional training for each attack separately, ii) testing on an unseen attack after training on others, and iii) incremental learning where the new attack is integrated into the existing model. This design allowed us to analyze not only the benefit of CIL in improving the detection of unseen attacks, but also its efficiency compared to full retraining. Importantly, we also analyzed that CIL substantially enhances performance on newly introduced classes while avoiding the catastrophic forgetting of previously learned attacks, which is an important point in incremental learning. In addition, our experiments demonstrate that incremental updates are four times faster than retraining from scratch, underscoring the practical viability of CIL for constrained IoT deployments. Together, we believe that these analyses establish CIL as an effective and efficient pathway to keep RPL-based intrusion detection systems resilient in the face of both known and emerging attacks. The remainder of this paper is organized as follows. First, we review the existing literature on RPL security. Next, we describe how class-incremental learning is implemented in RPL-based IoT networks. In Section \ref{sec:meth}, we then present the experimental evaluation in three stages: (i) detection of newly encountered attacks, (ii) assessment of catastrophic forgetting, and (iii) evaluation of computational performance. Finally, Section \ref{sec:conc} summarizes the key findings and outlines directions for future work.

\section{Literature Review} \label{sec:lite}
Most RPL defenses are designed for specific attack scenarios and lack mechanisms to adapt when new threats emerge (i.e., a static approach), whereas recent IoT intrusion detection system (IDS) studies highlight the potential of CIL to incorporate new attack classes without full retraining and to mitigate catastrophic forgetting. Empirically oriented works on IoT IDS show that CIL reduces retraining cost and latency for deployment while maintaining competence on previously learned behaviors, supporting its practicality for constrained RPL deployments \cite{CERASUOLO2025111228}.

Al Sawafi et al. \cite{2023Sawafi} proposed a hybrid deep learning (DL) approach combining supervised and semi-supervised models, achieving strong results on both known and novel attacks. While effective, their framework still assumes static retraining, leaving open the question of handling new attack classes incrementally. Foley et al. \cite{2020Foley} targeted combined objective function attacks in RPL using a multilayer perceptron (MLP) and reinforcement learning (RL), highlighting the complexity of compound threats. However, their work remains bound to fixed-class scenarios. Similarly, Almomani et al. \cite{2016Almomani} and Verma et al. emphasized dataset creation (WSN-DS, RPL-NIDDS17), underscoring the obsolescence of general benchmarks. Later Contiki-NG evaluations confirmed that ensemble methods, particularly XGBoost, consistently outperform other models \cite{10.1007/s11277-019-06485-w}. Yet, none of these studies examined adaptation when unseen RPL attacks arise.

Recent works also extend intrusion detection to edge devices. For example, an adaptive security framework for smart cameras introduced, tailored for resource-constrained environments \cite{10897877} . While showing the potential of resource-efficient DL, this line does not address catastrophic forgetting or incremental learning when new attack classes emerge. Transfer learning, a general category of incremental learning, is also used to bridge data scarcity and domain shift between heterogeneous IoT segments, and has been paired with federated learning in RPL settings. A representative example is an FTL-CID design that transfers a base model between server and edges in a heterogeneous RPL network to improve detection of existing and novel intrusions, illustrating knowledge reuse across topologies. Another study, FL-DSFA, addresses selective forwarding in RPL with federated learning, but scalability, energy efficiency, and real-time computation remain major concerns \cite{s24175834}. These limitations suggest that lightweight, adaptive learning is better suited for constrained IoT environments. Several studies also highlight transfer learning as a way to reuse knowledge across domains. Yang et al. \cite{9838780} and others \cite{10293947} used transfer learning with convolutional neural network (CNN) ensembles and particle swarm optimization (PSO) for IoT intrusion detection, while ByteStack-ID transformed packets into images for CNN-based classification \cite{10622596}. Cross-domain AI for O-RAN security also leveraged transfer learning and continuous training for evolving traffic \cite{10622345}. These efforts demonstrate knowledge reuse, but still stop short of evaluating incremental learning for RPL.

\section{The Proposed Approach}

\begin{figure}[t]
    \centering
    \includegraphics[width=\linewidth]{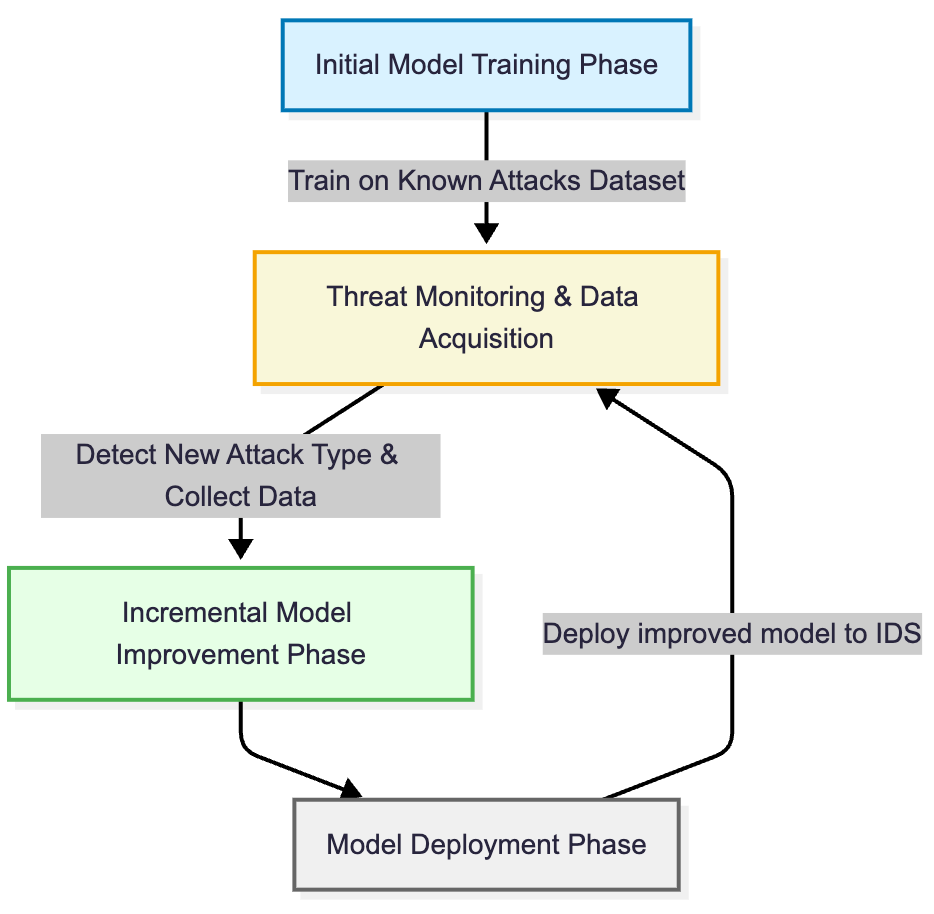}
    \caption{Workflow of the proposed class incremental learning (CIL) framework for RPL intrusion detection.}
    \label{fig:iot-rpl}
\end{figure}

The proposed adaptive RPL IDS framework, as shown in Fig. \ref{fig:iot-rpl}, follows a continuous process for detecting and adapting to evolving threats. It integrates a boosting-based model within a sequential fine-tuning pipeline. The system begins with initial training, where a base model is trained on a dataset containing known attack types and normal traffic. When a new attack is detected, the corresponding traffic data is provided to the system, triggering the incremental update phase. Instead of retraining from scratch, the framework employs a incremental learning mechanism: the pretrained boosting model is incrementally improved using only the new attack data, enabling it to learn additional classes without degrading its performance on previously learned ones.

The \textit{Initial Model Training Phase} trains baseline detectors on known attacks and normal traffic. The \textit{Threat Monitoring \& Data Acquisition} deploys IDS logs and RPL control/data features (e.g., DODAG information object rates, information solicitation rates, rank changes, parent switches, delivery ratios). Then, the \textit{Detect New Attack \& Collect Data} curates labeled samples when a previously unseen attack appears. In other words, we assumed the existence of a stable cluster corresponding to a previously unobserved behavior, following a similar approach to that proposed in \cite{9533187}. So, the framework declares a new attack class and creates a seed labeled set (active learning expands it using uncertainty and diversity sampling). After that, the \textit{Incremental Model Improvement Phase}, i.e., incremental learning, updates the deployed model to add the new class without full retraining. The objective of this phase is to update the IDS by incorporating newly discovered attack classes while avoiding \textit{catastrophic forgetting} of previously learned classes. Here, when a new class is identified, a new output unit is appended to the classifier layer. The backbone network is initialized from the previous model to preserve its feature representations.  A bounded exemplar buffer $\mathcal{M}$ is maintained to store a balanced subset of samples from previously known classes. To preserve knowledge of old classes, the new model  $f_{\theta}$ is trained to mimic the frozen teacher model $f_{\theta^{old}}$ using a distillation loss as follows. 
\begin{equation}
        \mathcal{L}_{KD} = T^2 \cdot \mathrm{KL}\!\left(
        \mathrm{softmax}\!\left(\frac{z^{old}}{T}\right) \,\bigg\|\, 
        \mathrm{softmax}\!\left(\frac{z^{new}}{T}\right) \right)    
\end{equation}
, where $z^{old}$ and $z^{new}$ denote the logits of the teacher and student models, respectively, and $T$ is the temperature \cite{Zhou_2024}.

The overall loss function combines cross-entropy for new and replayed exemplars, distillation loss, and a regularization term as follows. 
\begin{equation}
        \mathcal{L} = \mathcal{L}_{CE}(\text{new} \cup \mathcal{M}) 
        + \lambda \, \mathcal{L}_{KD} 
        + \gamma \, \mathcal{L}_{reg}
\end{equation}
, where $\mathcal{L}_{reg}$ (e.g., EWC or L2-SP \cite{Zhou_2024}) penalizes significant deviations in parameters deemed important for previously learned tasks. Finally, the \textit{Model Deployment} hot-swaps the improved model; the loop repeats as new threats emerge.

\begin{table*}[ht]
\centering
\caption{CIL benefit on unseen RPL attacks (F1 score comparison). R1: conventional training (upper bound), R2: unseen class, R3: after CIL. 
Recovery\% = $\frac{\text{F1}_{\text{R3}}-\text{F1}_{\text{R2}}}{\text{F1}_{\text{R1}}-\text{F1}_{\text{R2}}}\times100$. Best per-attack/model in \textbf{bold}.}
\label{tab:cil-main}
\begin{tabular}{ll
                S[table-format=1.2]
                S[table-format=1.2]
                S[table-format=+1.2]
                S[table-format=1.2]
                S[table-format=+1.2]
                S[table-format=3.0]}
\toprule
\multirow{2}{*}{Model} & \multirow{2}{*}{Attack}
& \multicolumn{1}{c}{R2: Unseen}
& \multicolumn{1}{c}{R3: CIL}
& \multicolumn{1}{c}{$\Delta$ (R3{-}R2)}
& \multicolumn{1}{c}{R1: Upper}
& \multicolumn{1}{c}{Gap (R1{-}R3)}
& \multicolumn{1}{c}{Recovery\%} \\
\cmidrule(lr){3-8}
 &  & {F1} & {F1} &  & {F1} &  & {(\%)} \\
\midrule
\multirow{3}{*}{XGBoost}
 & HF  & 0.9533	& 0.9924 &	\cellcolor{Gain} +0.0391 &	\bfseries 0.9934 & \cellcolor{Gap} + 0.0401 &	97.50623441 \\
 & DR  & 0.4615 & \bfseries 0.5333 & \cellcolor{Gain} +0.0718 & 0.5161 & \cellcolor{Gap} +0.0546 & 131.5018315 \\
 & VN & 0.6745 & 0.9463 & \cellcolor{Gain} +0.2718 & \bfseries 0.9532 & \cellcolor{Gap} +0.2787 & 97.52421959 \\
\midrule
\multirow{3}{*}{LightGBM}
 & HF  & 0.9449 & 0.9730 & \cellcolor{Gain} +0.0281 & \bfseries 0.9908 & \cellcolor{Gap} +0.0459 & 61.22 \\
 & DR  & \bfseries 0.6571 & 0.5872 & \cellcolor{Gain} -0.0699 & 0.5263 & \cellcolor{Gap} -0.1308 & 53.44 \\
 & VN & 0.6316 & 0.9401 & \cellcolor{Gain} +0.3085 & \bfseries 0.9493 & \cellcolor{Gap} +0.3177 & 97.104 \\
\midrule
\multirow{3}{*}{CatBoost}
 & HF  & 0.9291 & \bfseries 0.9908 & \cellcolor{Gain} +0.0617 & 0.9730 & \cellcolor{Gap} 0.0439 & 140.546697 \\
 & DR  & 0.4694 & \bfseries 0.5169 & \cellcolor{Gain} +0.0475 &  0.5055 & \cellcolor{Gap} +0.0361 & 131.578 \\
 & VN & 0.6467 & \bfseries 0.9469 & \cellcolor{Gain} +0.3002 & 0.9449 & \cellcolor{Gap} +0.2982 & 100.67 \\
\midrule
\multirow{3}{*}{DL}
 & HF  & 0.9220 & \bfseries 0.9543 & \cellcolor{Gain} +0.0323 & 0.9529 & \cellcolor{Gap} +0.0309 & 104.53 \\
 & DR  & 0.5106 & 0.5437 & \cellcolor{Gain} +0.0331 & \bfseries 0.6182 & \cellcolor{Gap} +0.1076 & 30.7620 \\
 & VN & 0.4495 & 0.9215 & \cellcolor{Gain} +0.4720 & \bfseries 0.9324 & \cellcolor{Gap} +0.4829 & 97.74 \\
\midrule
\multirow{3}{*}{GNN}
 & HF  & \bfseries 0.9359 & 0.9347 & \cellcolor{Gain} -0.0012 & 0.9086 & \cellcolor{Gap} -0.0273 & 4.395 \\
 & DR  & \bfseries 0.6412 & 0.6372 & \cellcolor{Gain} -0.0040 & 0.5824 & \cellcolor{Gap} -0.0588 & 6.802 \\
 & VN & 0.6747 & 0.9228 & \cellcolor{Gain} +0.2481 & \bfseries \bfseries 0.9304 & \cellcolor{Gap} 0.2557 & 97.027 \\
\midrule
\multicolumn{2}{r}{\emph{Mean across attacks}} 
& 0.7001 & 0.82274 & \cellcolor{Gain} 0.09964 & 0.8184933 & \cellcolor{Gap} +0.118359 & 83.4903293296 \\
\bottomrule
\end{tabular}
\begin{minipage}{0.94\linewidth}
\scriptsize \vspace{4pt}
HF: Hello Flood, DR: Decreased Rank, VN: Version Number. Shaded green shows gain ({$\Delta$}) and gives how much adding CIL helps when the third class is unseen at train time. Shaded red shows the remaining gap and shows how close CIL gets to the single-class upper bound.
\end{minipage}
\end{table*}

\section{Experiments} \label{sec:meth}

Experiments are designed to evaluate the performance of our framework across three distinct dimensions. The models selected for this study are representative of the most widely adopted machine learning techniques, chosen to provide a comprehensive view of how different architectures handle the challenges of incremental learning. The study includes three prominent boosting models, and a fully connected DNN is used as a representative of conventional DL architectures. Finally, a GNN with a GraphSAGE encoder \cite{10622177} is employed to evaluate the framework's performance on complex, relational network data. 

In this methodology, a binary classification task is performed by all models. Their prediction is always either malicious or normal traffic. Therefore, what is meant by a model ``learning a new attack type" is its ability to correctly classify instances of this novel attack as malicious, successfully integrating it into the existing understanding of malicious behavior. During the experiments, the size and best hyperparameters of each model are selected. For the boosting models, different hyperparameter values are tested to allow a fair comparison. These models use an $n_estimators$ range from $-1$ to $20$ and a $max_depth$ range from $2$ to $10$. A fixed $random_state$ of $3$ is applied to all models to make the results reproducible. The DL model uses the $Nadam$ optimizer, the $binary_crossentropy$ loss function, and the $AUC$ metric to measure performance in the binary classification task. In the GNN model, graph embeddings are used as input features. These embeddings are learned by maximizing the mutual information between positive and negative samples. All details can be found in our code repository \footnote{https://github.com/sumeyyebaas/Adaptive-Intrusion-Detection}.

For this study, network traffic data is generated within a simulated environment using Contiki-NG. The dataset comprised both normal traffic and traffic from three specific attack types: hello flood (HF), decreased rank (DR), and version number (VN) attacks. The raw data are divided into one-second intervals. For each interval, several metrics are calculated, including the number of packets, message types, and packet lengths. The same experimental procedure is applied to all three attack types, and the results from each are included in the final analysis.

\begin{figure*}[ht]
    \centering
    \includegraphics[width=\linewidth]{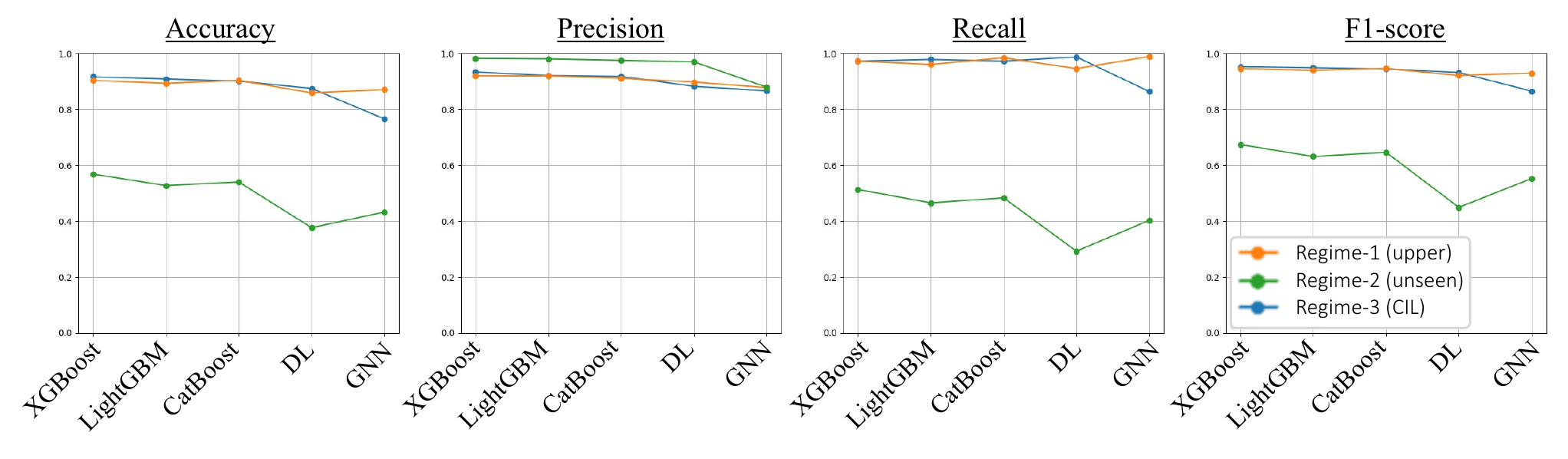}
    \caption{Comparison of model performance across different regimes when VN attack data are used as the test set. Accuracy, precision, recall, and F1-score are reported for five models (XGBoost, LightGBM, CatBoost, DL, and GNN). Regime-1: Models trained on VN and normal data, tested on VN data. Regime-2: Models trained on HF, DR, and normal data, tested on VN data (unseen attack scenario). Regime-3: Models initially trained on HF, DR, and normal data, then incrementally improved with VN data, and tested on VN data (incremental learning scenario).}
    \label{fig:test_scores}
\end{figure*}

\subsection{Experiment 1: New Attack Detection Evaluation}
The first experiment focus on how well the model detects previously unseen attack types. Here, we design three training regimes to reflect different operational scenarios. Regime 1 \textit{(R1 – Conventional Upper Bound)} represents the naturally best-case setting, where the model is trained on one specific attack type together with normal traffic and then tested on the same attack. This serves as an upper-bound reference since the model has full prior knowledge of the target class. Regime 2 \textit{(R2 – Unseen Class)} simulates the challenge of a novel intrusion, where the model is trained on two known attack types plus normal traffic and then evaluated on a third, previously unseen attack. Performance typically drops here, as the model has never been exposed to the new class during training. Regime 3 \textit{(R3 – Class Incremental Learning, CIL)} builds directly on R2, starting from the same model. We incrementally introduce the new attack class and improve the classifier without retraining from scratch.

To quantify the effect of this update, we also define several measures. CIL gain ($\Delta$) is the improvement (or sometimes deterioration) achieved by R3 over R2, i.e., how much the incremental update lifts detection performance on the new class, shown as
$\Delta = F1^{R3} - F1^{R2}$. The Gap refers to the difference between R1 (upper bound) and R3, indicating how far the incrementally updated model remains from the best possible performance, presented as
$\text{Gap} = F1^{R1} - F1^{R3}$. Finally, the Recovery measures the fraction of lost performance regained through CIL, computed as $\text{Recovery} = \frac{F1^{R3} - F1^{R2}}{F1^{R1} - F1^{R2}} \times 100\%$.



The primary research question here is whether it is effective to adapt an existing model to recognize new threats, instead of building a completely new model for each emerging attack. The results obtained with the VN test data are shown in Fig. \ref{fig:test_scores}. As expected, the model trained with other attack types is initially found to be unsuitable for the new attack type. In addition to the model's low success, it is observed to have a very high tendency to miss attacks, given its high precision value. It is an expected outcome that the model tends to label this new attack type as normal traffic because it cannot relate it to the other attacks. With the proposed incremental learning approach, when the models are retrained with the new attack data, a significant increase in model success is observed (see the blue line across plots). The high success in all metrics also confirmed its ability to truly distinguish attacks. The fact that it yielded very similar results to the model trained from scratch with only the new attack type is further evidence of its success in learning this new attack. Another conclusion learned from these results is that while boosting models achieved high success, DL and GNN models performed less successfully. The linear structure of the DL model and the overly complex structure of the GNN model's incompatibility with the data can be shown as a reason for this. The success of boosting models, which are frequently used in IDS studies \cite{10494996}, has also been confirmed in this study. 

Moreover, Table \ref{tab:cil-main} shows the increase in model performance achieved with incremental learning for other attack types. The contribution of the proposed model is observed to vary greatly depending on the type of attack. In cases where there is a decrease, the model's performance converged with that of a model trained from scratch with that attack type. For some examples, this convergence means achieving a huge difference. In other examples, the improved model is found to be more successful than the one trained from scratch. It is also clear that the similarity of the new attack type to previous attacks and the overall success of the models in the data are influential in these results. For the VN attack, it is observed that the proposed model always contributed to every model type. For the HF attack type, while the model trained with other attacks could already perform at a high level, it is seen that success could be further increased through improvement. For the DR attack, although there are examples where the CIL gain is negative, the performance of the improved model stands out compared to the model trained from scratch.

\subsection{Experiment 2: Catastrophic Forgetting Assessment}

The second experiment is dedicated to a critical challenge in incremental learning: \textit{catastrophic forgetting}. After the model has been updated with the new attack type, its performance is re-evaluated on the previously known attacks to confirm that adaptation does not compromise its ability to detect earlier threats. To conduct this assessment, a dedicated test set is held out from the previously known attack dataset. Models trained on the available training data are first evaluated on this held-out set to establish a baseline performance on the pre-existing attack types. After these models are incrementally updated with the new attack type, their test performance on the previously known attack types is measured once more. A significant decline in performance at this stage would have indicated catastrophic forgetting.

\begin{figure}
    \centering
    \includegraphics[width=\linewidth]{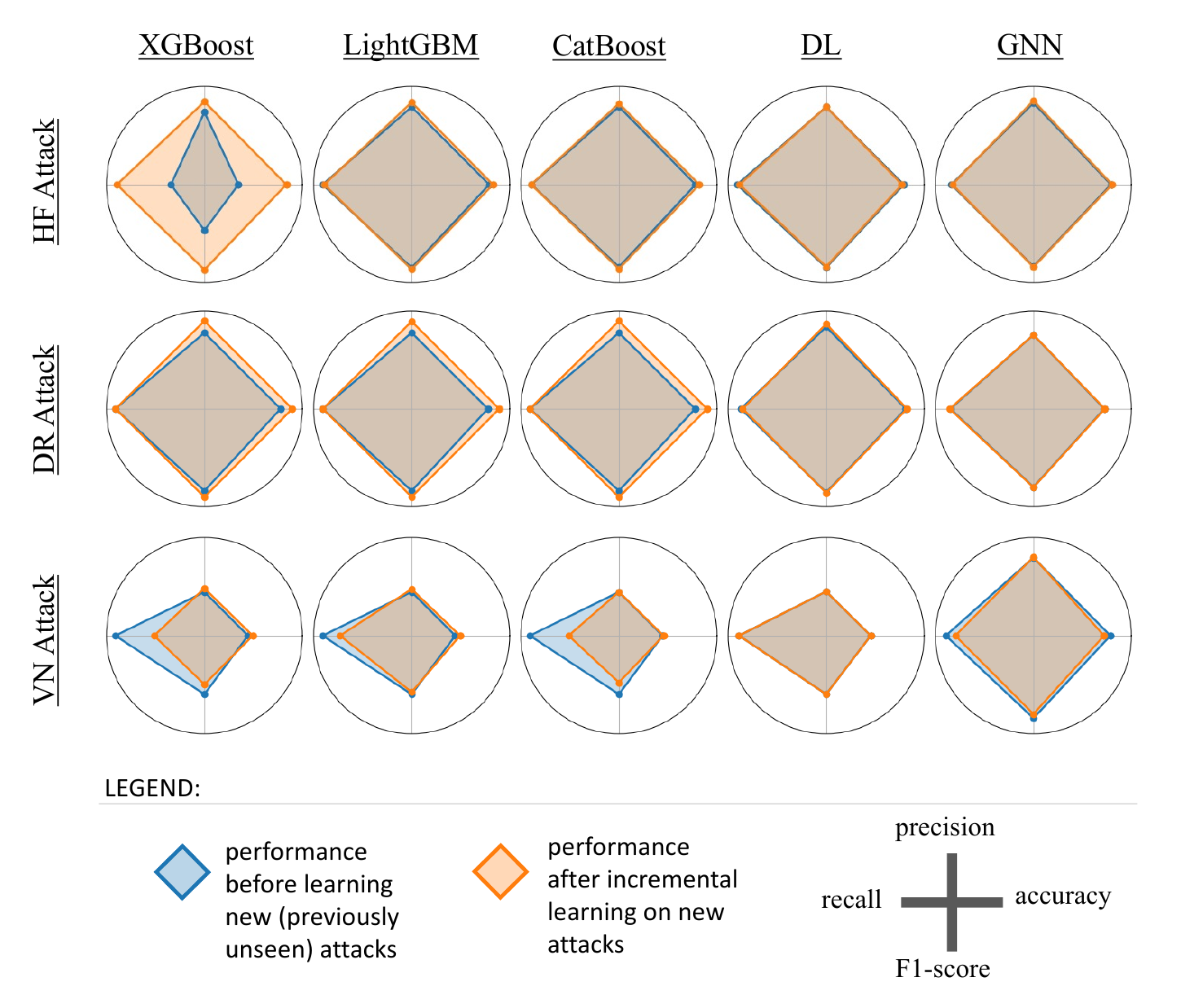}
    \caption{Change in the success of the model in pretrain data (blue: before previously known attacks, orange: after improving), right:accuracy, top:precision, left:recall, bottom:f1-score}
    \label{fig:pretrain_scores}
\end{figure}

The experimental results, presented in Fig. \ref{fig:pretrain_scores}, demonstrate that there is no significant reduction in the models' overall performance. On the contrary, the incorporation of certain new attack types even led to an improvement in the models' success on previously learned attacks. Especially in HF, boosting models improved their prediction performance in almost all metrics after improving. This may mean that learning HF attacks makes these models better understand other attack types. In the DR attack, it is observed that undesirable outcomes, such as high recall with low precision, are notably mitigated in the improved models. This is a common problem in imbalanced classification tasks, which means the model has so many false alarms. Using this kind of model leads to many unnecessary restrictions on non-attacker nodes. It is seen that, after improving models with DR data, boosting models overcome this problem on the previously known attacks.  For VN attacks, the XGBoost model has a drastic increase in all metrics after improvement. Considering its performance before improving, it can be said that improving the model on new attack types strengthens models even in the previous attacks. For all attack types, improving in DL and GNN models did not have a big impact. These models are not prone to forgetting previously known classes. However, comparing the improvements in the boosting models, it is clearly seen that they are more preferable for that task. 

Overall, experimental findings reveal that incremental learning not only strengthens the model’s ability to detect new attacks but can also boost its accuracy on previously learned ones.

\subsection{Experiment 3: Computational Performance Evaluation}

The third experiment addresses the practical demands of a real-world, operational security system by evaluating computational performance. The time-sensitive nature of an IDS requires that both the model's update process and its real-time prediction capabilities are highly efficient. 

\begin{figure}
    \centering
    \includegraphics[width=\linewidth]{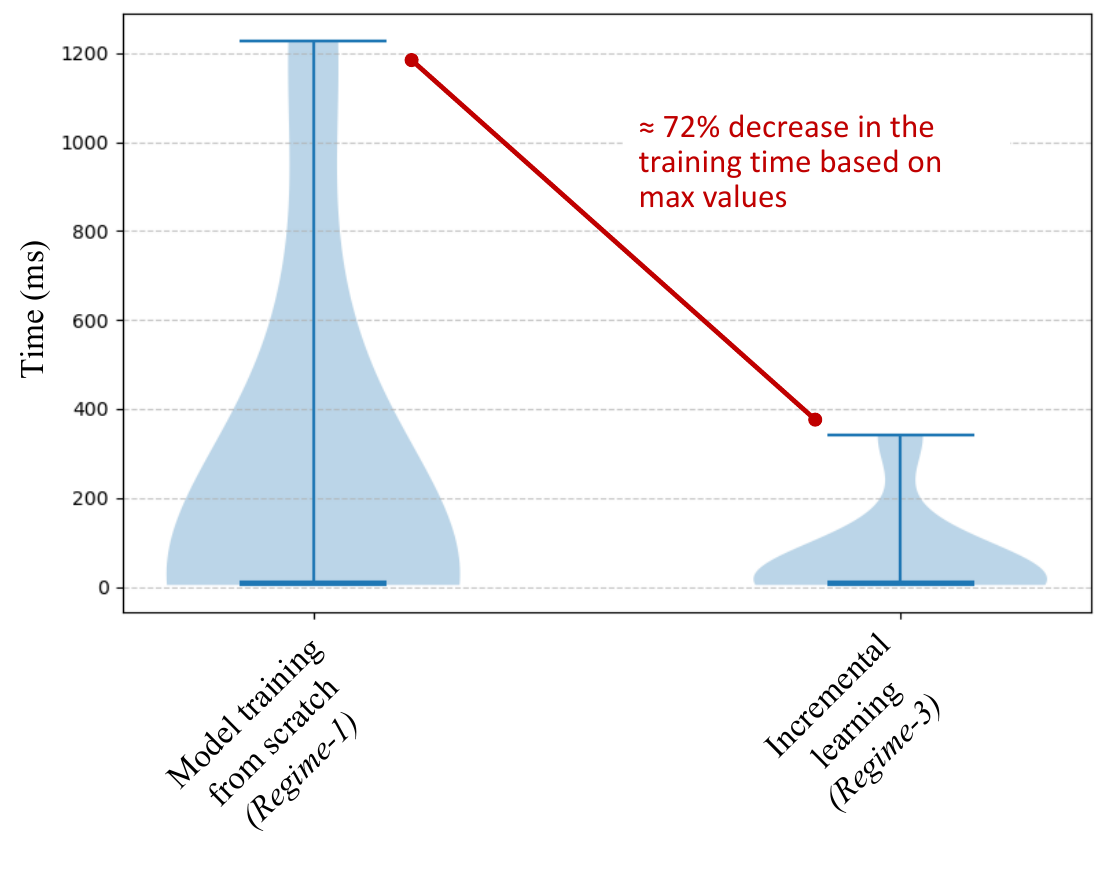}
    \caption{Comparison of the time required for incremental model updates and complete retraining from scratch.}
    \label{fig:time}
\end{figure}

To this end, a key objective is to quantify the time overhead of the incremental learning approach compared to training a model from scratch. Training times for each model while learning a new attack type were recorded, and their distributions are shown in Fig. \ref{fig:time}. The time difference depends on several factors, including the number of epochs, model type, and model size determined by hyperparameters. However, the main observation is that models previously trained on known attack types learn new data in a much shorter time. In fact, prior knowledge of different attacks enables models to learn new ones about 72\% faster when comparing maximum values. This result also highlights the time efficiency of the proposed approach and its strong suitability for the time-sensitive requirements of RPL networks.

\section{Conclusion} \label{sec:conc}

This study explored incremental learning as a practical and adaptive solution for intrusion detection in RPL-based IoT networks. Through extensive experiments on three common RPL attacks, i.e., hello flood, decreased rank, and version number, we demonstrated that incremental learning effectively restores detection performance on newly introduced attacks while preventing catastrophic forgetting of previously known ones. The approach also offers substantial time savings, with models learning new attacks up to 72\% faster than full retraining. Overall, the results highlight that incremental learning provides an efficient, resilient, and scalable foundation for securing evolving RPL networks against emerging IoT threats.
Future work will explore combined or concurrent RPL attacks, where multiple intrusion types occur at the same time. Another direction is to apply agentic AI, enabling systems that can monitor, decide, and adapt their own learning strategies. An agentic IDS could automatically detect changes in network behavior, update itself when needed, and maintain a balance between learning new attacks and retaining past knowledge.

\section*{Acknowledgements}
This research is supported by the Scientific and Technological Research Council of Turkey (TUBITAK) 1515 Frontier R\&D Laboratories Support Program for BTS Advanced AI Hub: BTS Autonomous Networks and Data Innovation Lab. project number 5239903, and the ITU Scientific Research Project Fund under grant number YESAP-2025-47304.
\bibliographystyle{IEEEtran}
\bibliography{mybibfile.bib}

\end{document}